\documentclass[a4paper]{article}
\usepackage{graphicx}
\usepackage{amsmath}
\usepackage{hyperref}
\usepackage{color}

\begin{document}

\begin{center}

{\bf \Large Clusterization, frustration and collectivity in random networks }\\[5mm]

{\large Anna Ma\'nka, Krzysztof Malarz and Krzysztof Ku{\l}akowski$^*$ }\\[3mm]

{\em

Faculty of Physics and Applied Computer Science,

AGH University of Science and Technology,

al. Mickiewicza 30, PL-30059 Krak\'ow, Euroland

}

\bigskip

$^*${\tt kulakowski@novell.ftj.agh.edu.pl}

\bigskip

\today

\end{center}

\begin{abstract}

We consider the random Erd{\H o}s--R\'enyi network with enhanced clusterization and Ising spins
$s=\pm 1$ at the network nodes. Mutually linked spins interact with energy $J$. Magnetic properties 
of the system as dependent on the clustering coefficient $C$ are investigated with the Monte Carlo 
heat bath algorithm. For $J>0$  the Curie temperature $T_c$ increases from 3.9 to 5.5 when $C$ increases 
from almost zero to 0.18. These results deviate only slightly from the mean field theory. 
For $J<0$ the spin-glass phase appears below $T_{SG}$; this temperature decreases with $C$, on the contrary
to the mean field calculations. The results are interpreted in terms of social systems.
\end{abstract}

\noindent

{\em PACS numbers:} 89.65.-s, 64.90.+b

\noindent

{\em Keywords:} random networks; phase transitions;

%% #####################################################################

\section{Introduction}

%% #####################################################################

Research on systems with random topology was initialized by Paul Flory in 1941 \cite{pf,ds0}, but they 
reappeared only about nine years ago \cite{wst} as a subject of great interdisciplinary
interest with applications in many sciences, from transport to biology \cite{r1,r2,r3}. Most of effort is concentrated 
on the topological structure of growing networks, where the scale-free degree distribution has been found. However, 
progress is achieved in the science of networks as a whole. 

In particular, networks provide a convenient basis of modelling social systems \cite{djw}. However, the structure of social networks remains debatable.
As it was discussed by Michael Schnegg \cite{schnegg}, the actual topology of a social network depends on social relations in the society.
Schnegg investigated six small-scale societies, mostly African ones. The degree distribution of three of them were found to be close to decreasing 
functions with some fluctuations, whereas the other three displayed a maximum. As the determinant what allows to differ between these
two groups, Schnegg proposed the amount of reciprocity; when it is large, the model degree distribution is close to the Poisson
function, known to appear in the classical Erd{\H o}s--R\'enyi random networks (\cite{r1,erd}. When the reciprocity is small, the 
system is supposed to be close to a scale-free network \cite{schnegg}. It seems that the topology of a social structure varies from one society to another, and various kinds of networks can be appropriate in different cases. We note only that, as remarked in Ref. \cite{new}, the 
clustering coefficient in the social networks is much larger, than in the Erd{\H o}s--R\'enyi networks.

On the other hand, the topology itself does not reflect the richness of behaviour of the 
social systems, and it is worthwhile to develop a theoretical description of these systems with more degrees of freedom. The 
simplest method is to decorate the network nodes $i=1,\dots,N$ with additional variables, as Ising spins $s_i=\pm 1$. 
These variables are not statistically independent. The system total energy is given as $E=-J/2\sum_{ij} s_is_j$, where 
the summation goes via all connected pairs of nodes/spins and $J$ is the so-called exchange integral. Two basic cases: 
ferromagnetic interaction ($J>0$) which prefers the same 
sign of neighboring spins, and antiferromagnetic interaction ($J<0$), which prefers their opposite signs, are to be discussed 
separately; for a recent review see \cite{dgm2}. In each case, the interaction competed with some noise, commonly 
modeled as the thermal noise. In numerous cases, a variation of the noise intensity leads to a phase transition, from the 
phase with ordered spins (low noise) to the phase of disordered spins (large noise). The amount of noise where the 
ordered phase disappears is comparable to the Curie temperature in ferromagnets. In fact, spin degrees of freedom ruled by
some stochastic evolution has been used many times to describe social processes, as for example the opinion dynamics 
\cite{sh}. 
In these works, ordering of spins can be treated as a demonstration of a collective behaviour despite of the presence 
of noise. The noise level is measured by parameter $T$ and usually called temperature.
Then the probability of move in the system states space from state `1' of energy $E_1$ to the state `2' of energy $E_2$ may be given
 by the probability $p_{1\to2}=[1+\exp(2\Delta E/T))]^{-1}$, where $\Delta E=E_2-E_1$. Such dynamic rule is termed as the heat--bath 
algorithm \cite{mich}.
When we deal with the social systems, a society can be modeled by the network and the Curie temperature ($T_c$) can be treated as 
a measure of an ability of a society to a collective action. This ability is expected to depend on the topology of 
interpersonal bonds. In terms of magnetism, the Curie temperature depends on the topology of the investigated network \cite{dgm}.

The case of antiferromagnetic interaction is of special interest for sociophysics. As it was argued only recently \cite{wss},
a dichotomous behaviour is found in some circumstances, as buying or selling \cite{min}. Another example is the dove or 
hawk strategy  
\cite{pds}. In this case the topology of the social network is particularly important because of the geometrical frustration effect,
which removes the unique and deep minimum of the magnetic energy. Instead, numerous local stable or metastable states
appear, and the structure of the set of these states remains unsolved \cite{byn}. We note in passing that the dynamics of the
magnetic behaviour of antiferromagnetic growing networks was found to be surprisingly rich \cite{mal}.

Here we are interested in the Erd{\H o}s--R\'enyi random network decorated with Ising spins, with ferromagnetic or 
antiferromagnetic interaction.
To enhance the clusterization coefficient, we apply the algorithm proposed by Holme and Kim
in Ref. \cite{hol} for the growing networks; it is easy to generalize it for the random networks. Other methods of network design 
have been proposed for example in Ref. \cite{maruda}. Our aim is to investigate, 
how the clustering influences the system behaviour. 

The Curie temperature for the random network was obtained already by Sumour et al. \cite{sum}. However, 
up to our knowledge the clusterization coefficient has not been varied in these networks.
Recently, similar problem for $J>0$ was also considered \cite{mac} in the hierarchical network, designed 
by Watts et al. \cite{wdn}.
In that network, the clustering was controlled by the probability of linking different clusters. Varying it, one moved from
the random network to a set of separate clusters. The obtained Curie temperature varied from $3.8J$ 
to zero in a step-like way, very sharply. Owing to algorithmic/numerical peculiarities of the hierarchical network, 
the size of the system
investigated in \cite{mac} was limited. Also, the clustering led the system to split into subnetworks, what does not 
occur in our case. In the antiferromagnetic case $J<0$, the uncorrelated network is believed to show the spin-glass phase
at low temperature \cite{rf1,rf2}. As all $J$'s are negative, there is no bond disorder; the frustration is introduced by 
the disorder of the network structure and it is purely geometrical \cite{gfr}.

%% #####################################################################

\section{The results}

%% #####################################################################

At the first stage of the simulation, a random Erd{\H o}s--R\'enyi network is constructed. The starting point is a set of 
$N=10^6$ nodes. A link is placed between each two nodes with the probability $p$. Next, the clustering is enhanced according 
to the receipt in Ref. \cite{hol}: nodes are selected with two or more neighbours, 
and a new link is placed between each pair of these neighbours with probability $r$ selected as to get the average 
degree $\langle k\rangle=4$. 

\begin{figure} 
\vspace{0.3cm} 
{\par\centering \resizebox*{10cm}{7cm}{\rotatebox{-90}{\includegraphics{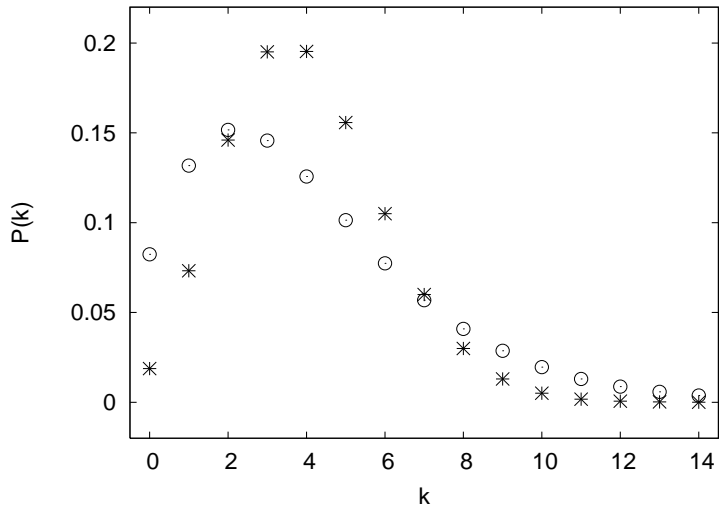}}} \par} 
\vspace{0.3cm} 
\caption{The degree distribution for the random network with $C\approx 0$  (stars) and $C=0.18$ (circles), for $N=10^6$ nodes.
The former is the Poisson distribution for $\langle k\rangle=4$ and $C\approx 0$; there, $\langle k^2\rangle=20.0$ and $z_2=16.0$.
For $C=0.18$ the simulation gives $\langle k^2\rangle=25.5$ and $z_2=21.5$.}
\label{fig-1}
\end{figure}

The clusterization coefficient $C$ is defined as the average over nodes $i=1,\dots,N$ of the local coefficient $C_i$, where
%% ---------------------------------------------------------------------

\begin{equation}
\label{eq-1}
C_i=\frac{2y_i}{k_i(k_i-1)}
\end{equation}
%% ---------------------------------------------------------------------
$k_i$ is the degree of $i$-th node {\it i.e.} the number of nodes linked to $i$, and $y_i$ is the actual number of 
links between these $k_i$ nodes. The maximal value of $C_i$ is one. In our system, the coefficient $C$ varies from almost 
zero (when $p=4/N$ and $r=0$) to $0.18$. This enhancement appears to influence the degree distribution.
As it is shown in Fig. \ref{fig-1}, the plot is Poissonian for small $C$, but deviates from this curve for $C=0.18$.

%% ---------------------------------------------------------------------

%% ---------------------------------------------------------------------

\begin{figure} 
\centering
\includegraphics[scale=0.8]{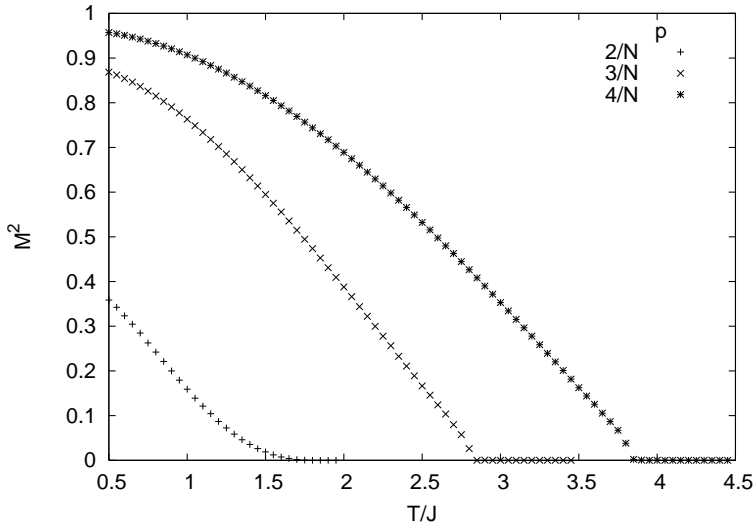}
\caption{Thermal dependence of the squared magnetization for $J>0$, $N=10^4$ nodes, $r=0$. The mean node degree 
$\langle k\rangle$ is 2 (left curve), 3 (middle) and 4 (right curve).
($N_{\text{iter}}=2\cdot 10^4$).}  
\label{fig-2}
\end{figure}

The magnetic behaviour of the system is also influenced by the clusterization. Let us start from the ferromagnetic case, 
where the exchange integral $J$ is set to +1.  We calculate the average magnetization $M=\sum_i s_i/N$ against temperature, $M(T)$,  for $r=0$. 
Initially, all spins are set to $+1$, and the time average is taken after $N_t$ time steps from another $N_t$ time steps. 
One timestep is equivalent to $N$ attempts to flip a spin. For the comparison with the results of \cite{sum}, the 
calculations for $\langle k\rangle=4$ are repeated for $\langle k\rangle=3$ and $\langle k\rangle=2$, $N=10^4$. The obtained curves for the squared magnetization are presented in Fig. \ref{fig-2}. As we see, the obtained Curie temperature
increases with $\langle k\rangle$. Also, the mean-field-character of the system is confirmed, as it was concluded in \cite{sum}, 
because the curves $M^2(T)$ are approximately linear. For $\langle k\rangle=2$, the value of the Curie temperature, $T_c/J\approx 1.7$ 
agrees with the one obtained in \cite{sum}.  For $\langle k\rangle=4$, the value of the Curie temperature $T_c/J\approx 3.8$ 
agrees with the result of \cite{mac} for the case of random network, where the homophily parameter $\alpha=-\ln 2$ 
(for the discussion of the interpretation of  $\alpha$ see Ref. \cite{wdn}). We note also that the analytical solution 
\cite{dgm} for the uncorrelated random networks with the Poissonian degree distribution 

%% ---------------------------------------------------------------------

\begin{equation}
\label{eq-2}
\frac{J}{T_c}=\frac{1}{2}\ln\left(\frac{\langle k^2\rangle}{\langle k^2\rangle-2\langle k\rangle}\right)
\end{equation}
%% ---------------------------------------------------------------------
gives $T_c/J=3.91$, 2.88 and 1.82 for $\langle k\rangle=4$, 3 and 2, respectively. The obtained values are slightly smaller. The 
differences are due mainly to the finite-size effect.

%% ---------------------------------------------------------------------

%% ---------------------------------------------------------------------

\begin{figure} 
\vspace{0.3cm} 
{\par\centering \resizebox*{10cm}{7cm}{\rotatebox{-90}{\includegraphics{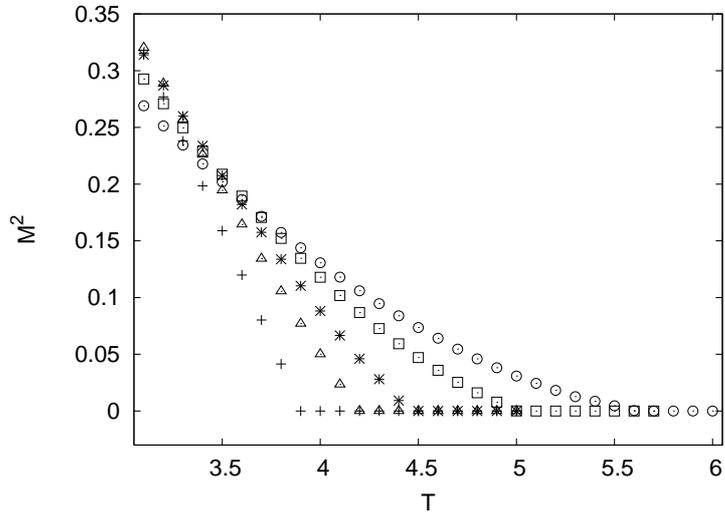}}} \par} 
\vspace{0.3cm} 
\caption{Thermal dependence of the squared magnetization for $J>0$, different values of $C$, and $N=10^5$ nodes. 
The mean node degree $\langle k\rangle =4$. The clustering coefficient $C\approx$ 0, 0.05, 0.09, 0.14 and 0.18 from left to right.}
\label{fig-3}
\end{figure}

\begin{figure} 
{\par\centering \resizebox*{10cm}{7cm}{\rotatebox{-90}{\includegraphics{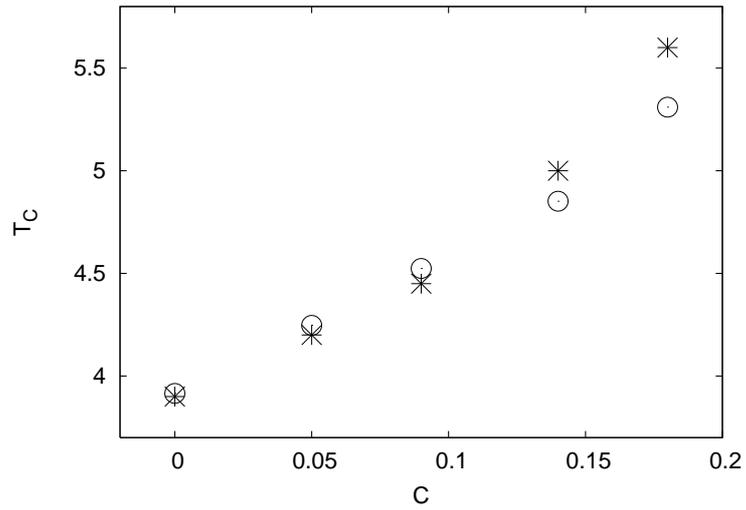}}} \par} 
\caption{The Curie temperature obtained from the simulation (crosses) and from Eq. \eqref{eq-2} or \eqref{eq-3} (circles).}  
\label{fig-4}
\end{figure}

These calculations were repeated again for  $\langle k\rangle=4$, $J>0$, $r>0$ and a larger network, i.e. $N=10^5$. The results shown 
in Fig. \ref{fig-3} indicate, that the Curie temperature increases with the clustering coefficient $C$. This increase does follow
from the increase of the second moment $\langle k^2\rangle$ of the degree distribution. As shown in Fig. \ref{fig-4}, the theoretical 
estimation of Eq. \eqref{eq-2} differs only slightly from the observed value of $T_c$. The same is true for the more general 
formula \cite{dgm2}
%% ---------------------------------------------------------------------

\begin{equation}
\label{eq-3}
\frac{J}{T_c}=\frac{1}{2}\ln\left(\frac{z_2+z_1}{z_2-z_1}\right)
\end{equation}
%% ---------------------------------------------------------------------
where $z_1=\langle k\rangle=4$, and $z_2$ is the number of second neighbours. Both equations 2 and 3 give the same theoretical values. 
Indeed, our numerical calculations confirm the relation $z_2=<k(k-1)>$.

\begin{figure} 
{\par\centering \resizebox*{10cm}{7cm}{\rotatebox{-90}{\includegraphics{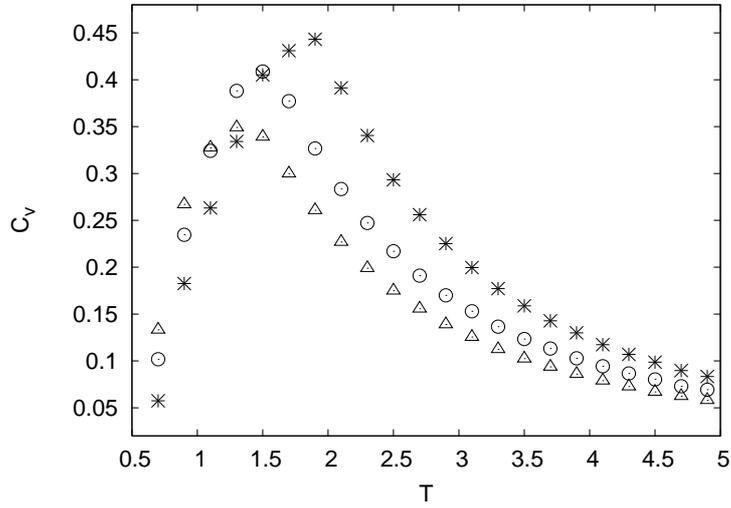}}} \par} 
\caption{The magnetic specific heat $C_V(T)$ for the case $J<0$, where $C$=0, 0.09 and 0.18 (crosses, circles and triangles, respectively).}  
\label{fig-5}
\end{figure}

In the antiferromagnetic case ($J<0$) the disordered topology of the random network does not allow to investigate the 
staggered magnetization as the order parameter. We made an attempt to calculate the magnetic specific heat $C_V$ against 
temperature. For the lattice antiferromagnets, it is the maximum of $C_V$ at given temperature what marks an existence 
of the phase transition. Indeed, we found an indication of such a transition for the case $r=0$, where the clusterization 
coefficient vanishes. Slightly below the maximum, $C_V$ obtained from the derivative of energy with respect to $T$ starts to
differ from $C_V$ calculated from the variance of energy. This split is due to the lack of thermal equilibrium in the spin-glass phase. Still,
the position of the maximum can be determined. In Fig. \ref{fig-5}, only the result from the thermal derivative of energy is shown.
These calculations are preceded by $N_t=10^4$ time steps, till the time dependence of energy reaches the plateau; this was checked to be 
long enough if $T>0.7$. 

\begin{figure} 
{\par\centering \resizebox*{10cm}{7cm}{\rotatebox{-90}{\includegraphics{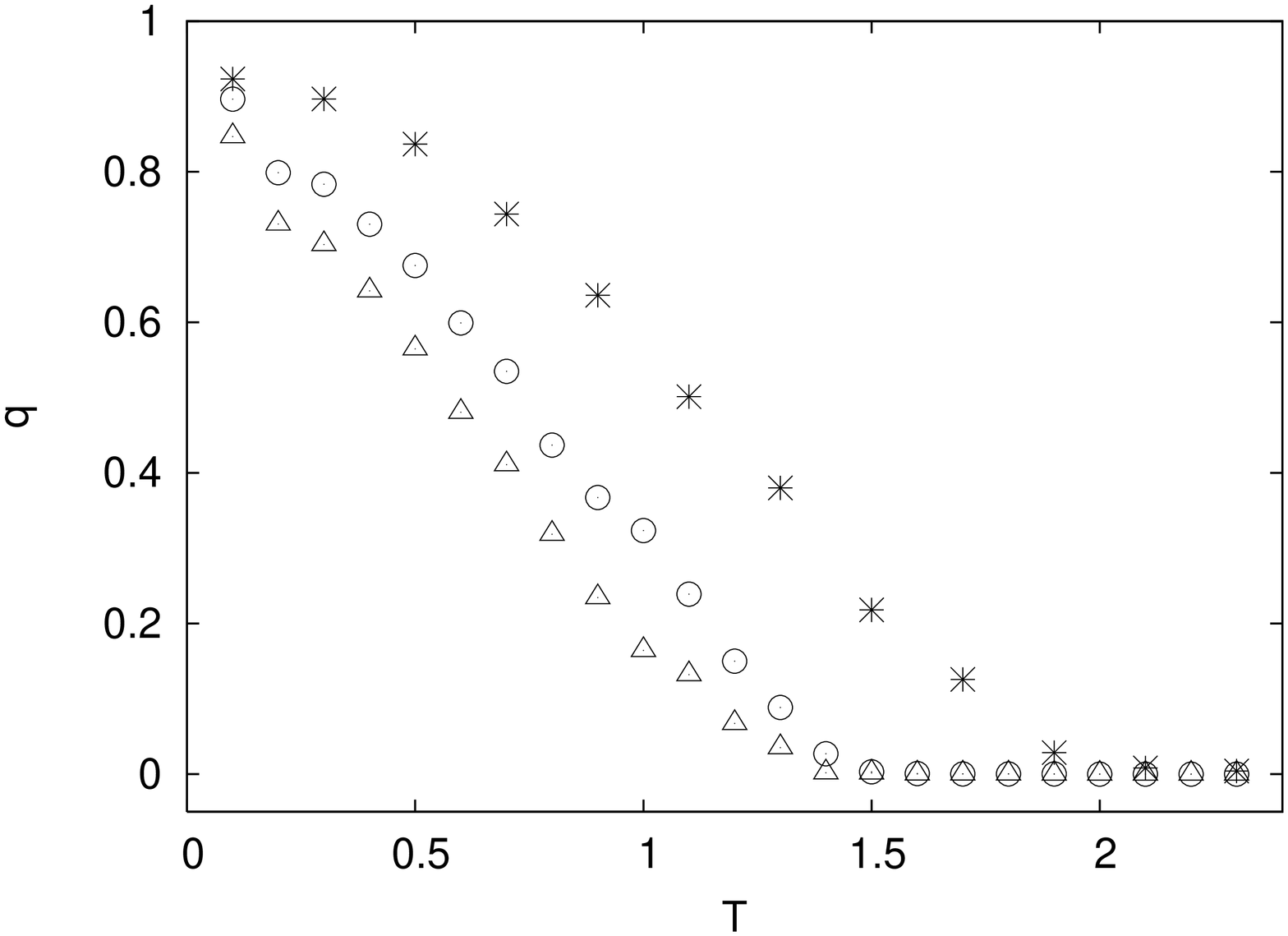}}} \par} 
\caption{The Edwards-Anderson order parameter $q(T)$ for the case $J<0$, where $C$=0, 0.09 and 0.18 (crosses, circles and triangles, respectively).}  
\label{fig-6}
\end{figure}

The plots shown in Fig. \ref{fig-5} show that the transition temperature $T_{SG}$ decreases with the clustering coefficient $C$. This result
is confirmed by the calculations of the Edwards-Anderson spin-glass order parameter $q$ \cite{ea,kraw}

\begin{equation}
\label{eq-4}
q=\frac{1}{N}\sum_i \Big(\frac{1}{\tau}\sum_{t=1}^\tau s_i(t)\Big)^2
\end{equation}
%% ---------------------------------------------------------------------
where $s_i(t)$ is $i$-th spin at time $t$. The obtained plots for $q(T)$ for various $C$ are shown in Fig. \ref{fig-6}. Here also, the 
temperature where $q$ vanishes decreases from $1.9$ for $C=0$ to about 1.4 for larger values of $C$.

%% ---------------------------------------------------------------------
%% ---------------------------------------------------------------------

%% ---------------------------------------------------------------------
%% ---------------------------------------------------------------------

%% ---------------------------------------------------------------------
%% ---------------------------------------------------------------------

%% ---------------------------------------------------------------------
%% ---------------------------------------------------------------------

%% #####################################################################
\section{Discussion}
%% #####################################################################

New results obtained for the ferromagnetic case indicate that the enhancement of the clusterization by 
additional links does not invalidate the assumption of the lack of correlations, which is usually made in the 
theory \cite{dgm} of random networks. However, for $C=0.14$ and above this value, the 
loops introduced to the system slightly enhance the Curie temperature, as shown in Fig. 4. 

For the antiferromagnetic case, the situation is more complex. For negligible clusterization, the maximum 
of the specific heat indicates the phase transition between the paramagnetic and the spin-glass phase. The transition 
temperature $T_{SG}$ rougly accords with the mean field result $T_{SG}=1.91$ from Eq.5. 

\begin{equation}
\label{eq-5}
\frac{z_1}{z_2}=\tanh^2(\frac{1}{T_{SG}})
\end{equation}
which follows from Eq. 92 in Ref. \cite{dgm2} and from our delta-like distribution $\rho (J)=\delta(J+1)$ of the 
exchange integral $J$. However, the theoretical values of $T_{SG}$ increase with $C$ to 1.92 for $C=0.09$ and to 2.17 for $C=0.18$. 
Our numerical results on $C_V$ and $q$ indicate, that actually  $T_{SG}$ decreases with $C$. 

The topology of the system allows to treat it as a disordered version of the Archimedean lattice, discussed in Ref. \cite{kraw}.
There, the triangles of spins were connected as to form the periodic structure. The ground state degeneracy of this Archimedean lattice
increases exponentially with the system size. The same rule applies to our disordered lattice if the density of the triangles is 
large enough.

As discussed in Ref. \cite{dgm2}, there are links from the problem of spin-glass in disordered networks to some interested NP-complete 
problems as MAX-CUT or the satisfiability. It seems that some counterparts of our system can be found also in the game theory, and in particular in the network congestion games \cite{ncg}; for a recent review of games on networks see Ref. \cite{games}. To maintain our social interpretation, 
we can state that the case of a ferromagnet the smooth increase of the Curie 
temperature with the coefficient $C$ marks the fact that the ordered phase is less fragile if the system has the clustered 
structure.  Considering an antiferromagnet we take into account that effects
of social interactions are sometimes incoherent, that the society can be polarized in some way and that it is possible that
increasing the density of social ties can be in conflict with an ability to find best solutions to everyday problems.
In this sense, geometrical frustration combined with noise can reproduce some social phenomena which disable a coherent 
social action. A next step to reflect features of the society should be to consider a distribution of positive and negative 
interaction of different sizes. This, however, requires much larger conceptual and numerical effort.  

\bigskip

\section*{Acknowledgements}
The authors are grateful to Dietrich Stauffer for helpful suggestions and critics. The calculations were performed in the ACK Cyfronet, Cracow, grants No. MNiSW /SGI3700 /AGH /030/ 2007 and  MNiSW /IBM BC HS21 /AGH /030 /2007.

\end{document}